\documentstyle[12pt,aps,prc,epsfig,preprint]{revtex}
\tightenlines
\begin{document}

\begin{titlepage}
\title{\vspace*{5mm}\bf
\Large Density dependence of the $s$-wave repulsion in pionic atoms}
\vspace{4pt}

\author{  E.~Friedman  \\
{\it Racah Institute of Physics, The Hebrew University, Jerusalem 91904,
Israel\\}}

\vspace{4pt}
\maketitle

\begin{abstract}

Several mechanisms of density dependence of the $s$-wave repulsion
in pionic atoms, beyond the conventional model, are tested by
parameter fits to a large (106 points) set of data from $^{16}$O to
$^{238}$U, including `deeply bound' states in $^{205}$Pb.
Special attention is paid to the proper choice of nuclear density
distributions. 
A density-dependent isovector scattering amplitude
suggested recently by Weise to result from a density dependence of the
pion decay constant is introduced and found to account for  most of 
the so-called anomalous repulsion. The presence of such an effect
might indicate  partial chiral symmetry restoration in dense matter.
The anomalous repulsion is fully accounted for when an additional relativistic
impulse approximation term is included in the potential.

$PACS$: 13.75-n; 13.75.Gx; 25.80.Hp
\newline
\newline \newline
Corresponding author: E. Friedman, \newline
Tel: +972 2 658 4667,
FAX: +972 2 658 6347, \newline
E mail: elifried@vms.huji.ac.il

\end{abstract}

\centerline{\today}
\end{titlepage}

\section{Introduction}
\label{sec:int}

Strong interaction effects in pionic atoms have been studied for
several decades both experimentally and theoretically \cite{BFG97}.
A theoretically-motivated phenomenological pion-nucleus potential \cite{EEr66}
has been quite successful in reproducing the experimental values of
strong interaction level shifts and widths throughout the periodic
table. This potential has a local part which is effective over the whole
of the nuclear volume and a $p$-wave part which is effective mostly in
the surface region of the nucleus.
 Both components have terms linear 
in the nuclear densities which
are closely related to the free pion-nucleon interaction, and quadratic
(complex) terms which originate from pion absorption on pairs of
nucleons. The latter are mostly determined empirically from fits to data
whereas the former, although determined empirically, may also be
calculated from the pion-nucleon interaction in free space.
The $p$-wave component of the potential turns out to be fairly well
understood, but the $s$-wave part of the potential had turned out
too repulsive compared with expectations \cite{OTK84,GOS88,OGN95}.

The renewed interest in pionic atoms in general and in the $s$-wave
part of the potential in particular stems from two recent
developments. The first is the experimental observation of `deeply
bound' pionic atom states in the (d,$^3$He) reaction 
\cite{YHI96,GGK00} the existence of which was predicted a decade earlier
\cite{FSo85,TYa88,THY89}. The second is the very accurate measurement
of the shift and width of the 1$s$ level in pionic
hydrogen \cite{SBG01} which leads to precise values of the $s$-wave
scattering lengths.

Very recent attempts to explain the above mentioned $s$-wave repulsion
in terms of a density dependence of the pion decay constant were
made by Weise and Kaiser \cite{Wei01,KWe01}. The proposed mechanism
was implemented in fits to a large set of pionic atom data \cite{Fri02}
and indeed found to remove most of the `anomaly' in the $s$-wave term.
A relativistic impulse approximation (RIA) term that was proposed
earlier following Birbrair \cite{BFG83,GLM91,BGr91,GJF92,BGL92} 
was also shown \cite{GJF92,Fri02}
 to remove part of the anomaly. Combining the two effects
 \cite{Fri02} removed the anomaly completely.

The present work is an
extention of Ref. \cite{Fri02} in several respects. First, the data
base of the present work contains 106 data points compared to 60 points
in the earlier work. The additional data are mostly from the work
of the Amsterdam group 
(see \cite{KLT90} and references therein)
which includes several sequences of
isotopes, of particular importance in the present context where most
of the effect is due to an isovector term (see below). The second
difference compared to the earlier work is in the nuclear density
distributions, where in addition to `macroscopic' densities \cite{BFG97}
we have used `single particle' densities constrained by results of
relativistic mean field calculations. The third difference is that in
the present work more flexibility was allowed in the RIA model and
in the $\chi ^2$ fits.
The general context of the present work is the use of several models
for the pion-nucleus interaction, each resulting in a well-defined 
functional of the local nuclear densities. Fits to a large
set of experimental data provide the parameters of these various
functionals.

The paper is organized as follows. Section \ref{sec:potl} describes the
pion-nucleus potential, including discussion of the nucleon
densities. Section \ref{sec:data} summarizes the data base and the
fit procedures. The results are given in Section \ref{sec:res}
and Section \ref{sec:sum} is a summary. 

\section{Theoretical background} \label{sec:potl}

The interaction of pions at threshold with the nucleus
is described by the Klein-Gordon (KG)
 equation of the form:

\begin{equation}\label{equ:KG1}
\left[ \nabla^2  - 2{\mu}(B+V_{{\rm opt}} + V_c) + (V_c+B)^2\right] \psi = 0~~ ~~
(\hbar = c = 1)
\end{equation}
where $\mu$ is the pion-nucleus reduced mass,
$B$ is the complex binding energy
 and $V_c$ is the finite-size
Coulomb interaction of the pion with the nucleus, including
vacuum-polarization terms.
Equation (\ref{equ:KG1}) assumes that 
the strong interaction potential $V_{{\rm opt}}$ behaves 
as a Lorentz scalar. The potential is usually taken as suggested by 
Kisslinger \cite{Kis55} and modified by Ericson and Ericson \cite{EEr66}
to include absorption of pions on pairs of nucleons. The form used in the
present work is

\begin{equation} \label{equ:EE1}
2\mu V_{{\rm opt}}(r) = q(r) + \vec \nabla \cdot \alpha(r) \vec \nabla
\end{equation}
with
\begin{eqnarray} \label{equ:EE1s}
q(r) & = & -4\pi(1+\frac{\mu}{M})\{{\bar b_0}(r)[\rho_n(r)+\rho_p(r)]
  +b_1[\rho_n(r)-\rho_p(r)] \} \nonumber \\
 & &  -4\pi(1+\frac{\mu}{2M})4B_0\rho_n(r) \rho_p(r) ,
\end{eqnarray}

\begin{equation} \label{equ:LL2}
\alpha (r) = \frac{\alpha _1(r)}{1+\frac{1}{3} \xi \alpha _1(r)}
 + \alpha _2(r) ,
\end{equation}
\noindent
where

\begin{equation} \label{equ:alp1}
\alpha _1(r) = 4\pi (1+\frac{\mu}{M})^{-1} \{c_0[\rho _n(r)
  +\rho _p(r)] +  c_1[\rho _n(r)-\rho _p(r)] \} ,
\end{equation}

\begin{equation} \label{equ:alp2}
\alpha _2(r) = 4\pi (1+\frac{\mu}{2M})^{-1} 4C_0\rho _n(r) \rho _p(r).
\end{equation}

\noindent
In these expressions $\rho_n$ and $\rho_p$ are the neutron and proton density
distributions normalized to the number of neutrons $N$ and number
of protons $Z$, respectively, 
and $M$ is the mass of the nucleon.
In this potential $q(r)$ is referred to
as the $s$-wave potential term and $\alpha(r)$ is referred to
as the $p$-wave potential term.
The function ${\bar b_0}(r)$ is given in terms of the {\it local} Fermi
momentum $k_{\rm F}$

\begin{equation} \label{equ:b0b}
{\bar b_0}(r) = b_0 - \frac{3}{2\pi}(b_0^2+2b_1^2)k_{\rm F}(r),
\end{equation}
where $b_0$ and $b_1$ are minus the pion-nucleon isoscalar
and isovector scattering lengths, respectively. 
The  coefficients $c_0$ and
$c_1$ are the pion-nucleon
isoscalar and isovector $p$-wave scattering volumes,
respectively. The parameters $B_0$ and $C_0$ represent $s$-wave and $p$-wave
absorptions, respectively, and as such they have imaginary parts.
Dispersive real parts are found
to play an important role in pionic atom potentials.
The parameter $\xi$ is the usual 
Ericson-Ericson Lorentz-Lorenz coefficient (EELL) \cite{EEr66}.
The terms with $4\rho _n \rho _p$ were originally written as 
$(\rho _n+\rho _p)^2$ (see Ref. \cite{EEr66}), but the results hardly
depend on which form is used. An additional relatively small term,
known as the `angle-transformation' term (see Eq.(24) of \cite{BFG97}),
is also included. The local $k_F (r)$ was taken from the Fermi gas model.
There is no risk of using this model for extremely low densities
because the major contributions to strong interaction effects come
from regions near the half-density point, regardless of where the
peak of the atomic wavefunction is. The position of that peak
relative to the nuclear surface determines only the scale of the effects.

The potential as described above had been used extensively to analyze
pionic atom data \cite{BFG97,KLT90}. It will be referred to in the following
as the conventional potential (C). Very good fits to the data are obtained
with this potential but the combined repulsion due to the 
resulting values of the parameters $b_1$
and Re$B_0$ is about twice as large as is expected from the very well
known values of the free pion-nucleon scattering lengths \cite{SBG01} and
from the well-determined parameter Im$B_0$ \cite{BFG97,KLT90}.
For that reason the $s$-wave interaction in the nuclear medium continued to
be a topic of interest. Very recently Weise \cite{Wei01} showed that if
the pion decay constant in nuclear matter $f_\pi^*$ is given, in leading
order, as a function of the local density $\rho$

\begin{equation} \label{eq:fpi2}
f_\pi ^{*2} = f_\pi ^2 - \frac{\sigma _N}{m_\pi ^2} \rho
\end{equation}
where  $f_\pi$ is the decay constant of the
pion in free space and $\sigma_N$ is the pion-nucleon sigma term,
then the $s$-wave scattering amplitude becomes a function of the local
density  as follows

\begin{equation}\label{equ:ddb1}
b_1(\rho) = \frac{b_1(0)}{1-2.3\rho}
\end{equation}
for $\sigma _N$=50 MeV and
with $\rho$ in units of fm$^{-3}$.
Note that expanding this expression in powers of the density leads
naturally to a repulsive $\rho ^2$ term.
 The parameter $b_1$(0) refers
to the experimental free pion-nucleon interaction at threshold \cite{SBG01}.
Introducing this function of the local density into the above potential,
in Eq.(\ref{equ:EE1s}) and (\ref{equ:b0b}), leads to the `W' potential of
Ref. \cite{Fri02}. If $b_1$(0) is taken as the first order chiral
perturbation result of Weinberg, then a third order correction 
of about 15\% brings it in
line with the experimental value of \cite{SBG01}. In that case there is
an additional density dependence in Eq.(\ref{equ:ddb1}) which, however, does 
not affect the results for pionic atoms beyond what is obtained when
Eq.(\ref{equ:ddb1}) is used with the empirical pion-nucleon value. We
therefore refer to the experimental free pion-nucleon value throughout, denoting
this potential by `W'.

At this point it is useful to state clearly the philosophy behind the
present work and the interplay between empirical quantities and more
fundamental quantities. The potential of Eq.(\ref{equ:EE1}) is said
to respect the low density limit when the coefficients of the linear
terms ($b_0, b_1, c_0, c_1$) are those obtained from the free pion-nucleon
interaction and this can be easily achieved (see below) for the $p$-wave part
of the potential. However, this is not the case with the parameters
$b_0$ and $b_1$, with the latter found to be too repulsive. The coefficient
Re$B_0$, although purely empirical, is also somewhat constrained
theoretically,  e.g. it
was shown \cite{GOS88} that it is very unlikely to be repulsive and
larger in absolute value than Im$B_0$. In practice the empirical values
of Re$B_0$ are at variance with this expectation, and together with the
repulsive $b_1$ lead to the so-called `anomalous' repulsion.
Because the $b_0$ parameter is very small
Weise \cite{Wei01} considered medium effects {\it only} on the parameter
$b_1$ suggesting that the empirical extra repulsion compared to the free
pion-nucleon interaction is due to medium modification of the pion
decay constant. In Ref.\cite{Fri02} and in the present work
we check that approach by large scale fits to data, while
treating $B_0$ (and $C_0$) as {\it purely phenomenological}. For that reason
no additional effects due to medium modification of the pion decay constant
are considered in the present work. Note, however, that  a recent
preprint \cite{GNO02} reports failure when additional medium modification
effects are considered.

Among several previous attempts to account for the 
anomalous $s$-wave repulsion 
a relativistic impulse approximation  approach 
showed \cite{BFG83,GLM91,BGr91,GJF92,BGL92}
 that  a specific version of the RIA
was able to provide a significant part of the anomalous repulsion
through the modification of the {\it nucleon} mass in the nuclear medium.
This version of the RIA was considered as an additional 
option in \cite{Fri02} and is included in
the present work.
The additional term of the potential, to be added to
Eq.(\ref{equ:EE1}), is

\begin{equation} \label{equ:Bir0}
2\mu \Delta V_{{\rm opt}}(r) = -4 \pi \frac {m}{M} d_0 (\frac {M^2}
  {M^2 (\rho )} -1) \rho _k (r)
\end{equation}
where $M(\rho )$ is the effective mass of the nucleon in the medium,
$\rho $ is the total density and
$\rho _k (r)$ is the squared nucleon momentum distribution given
in the local density approximation by

\begin{equation} \label{equ:Fer}
\rho _k (r) = \frac{3}{5} (\frac{3}{2} \pi ^2)^{2/3} \rho ^{5/3} (r).
\end{equation}
The  coefficient $d_0 = -0.190 m_\pi ^{-3}$ originates in
the spin-dependent interaction \cite{EEr66}. For the effective nucleon mass
the following parameterization was used \cite{BGr91}

\begin{equation}\label{equ:Bir}
\frac{M(\rho )}{M(0)} = \frac{1}{1 + a \rho}.
\end{equation}
\noindent
With $a$=2.7 fm$^3$, we find $M(\rho)/M(0)=0.7$ for
the nucleon mass ratio at normal nuclear
density. With $a$=1.56 fm$^3$ this ratio is 0.8. Both values have been used in 
the present work. Note, however, that this version of the RIA is
not unique \cite{GJF92}.

The nuclear densities $\rho_p$ and $\rho_n$ are essential ingredients of
the pion-nucleus potential, and have been discussed at some length 
in \cite{BFG97}. Two questions are relevant in connection with nuclear
densities: (i)the model used to generate the densities and (ii) the
radial extent of the neutron densities. The 
root mean square (rms) radii of the proton densities are obtained
from the experimentally determined charge distributions \cite{FBH95} by
unfolding the finite size of the proton.
In the present work we have used both the macroscopic (MAC) densities
and the single particle (SP) densities discussed in \cite{BFG97}. 
In previous analyses \cite{KLT90,BFG97,Fri02} the rms
radii of the neutron densities ($r_n$) were assumed equal to the corresponding
rms radii for the protons ($r_p$) in the case of light ($N=Z$) nuclei, and were
taken to be slightly larger than the corresponding radii for the protons
for $N>Z$ nuclei. Here we have used those previous values and in addition  
we used rms radii for the neutron densities as obtained 
from recent relativistic
mean field (RMF) predictions \cite{LRR99} for the {\it differences} $r_n-r_p$.
Moderate sensitivity to values of $r_n$ for {\it given} sets of potential
parameters is observed, e.g. 100 and 50 keV for the binding energies of
the 1$s$ and 2$p$ states in $^{205}$Pb, respectively, for a decrease of $r_n$
of 0.1 fm. However, once best fits to the data have been made,
the sensitivity of the results to the precise values of $r_n$ is small.
Nevertheless, attempting to use neutron densities with $r_n=r_p$ resulted in
major deterioration in the quality of fits to the data.
Since the deeply bound pionic atom states of $^{205}$Pb \cite{GGG02}
may arise some interest we quote the rms radii used, namely 5.45 fm for the
protons and 5.63 or 5.71 fm for the neutrons in the two sets of fits
performed for each type of density as described above.

Finally we remark that the strong interaction shifts are 
always taken relative
to the corresponding electromagnetic values calculated from the {\it finite
size} charge of the nucleus and including vacuum polarization terms. This is
the conventional definition and actually the only one meaningful for 1$s$
states in heavy nuclei ($Z>137/2$).

\section{Data base and fit procedures}  \label{sec:data}

The data base for the present work of 106 data points was made  
by adding 46 points to the 60 points used in Ref. \cite{Fri02}. Those 
60 points were the 54 points of Ref. \cite{BFG97} to which were added
the shift and width of the 4$f$ level in 
$^{208}$Pb \cite{KLT90} and the recently
obtained results \cite{GGG02} for the `deeply bound' 
1$s$ and 2$p$ states in 
pionic atoms of $^{205}$Pb. In the case of the deeply bound states the
shifts, relative to the finite size and vacuum polarization values
discussed in the previous section, were simply obtained from the experimental
binding energies \cite{GGG02}. The shift of the 4$f$ level 
as well as the 23 additional shifts \cite{KLT90}
were transformed to our normal basis of the finite size plus vacuum 
polarization reference using the parameters of the charge distributions
listed in Ref. \cite{KLT90}.

The parameters of the pion-nucleus potential were varied in $\chi ^2$
minimization, where $\chi ^2$ was defined in the usual way. With so
many data points it was possible to vary simultaneously all 9
parameters of the potential, but obviously some were determined better
than others and, moreover,  correlations exist between 
some of the parameters.
Parameter values in the 9-dimensional space were chosen according to
the following criteria, in descending order of importance:

\begin{enumerate}
\item The lowest possible total $\chi^2$ was required to within the natural
  unit for this problem which is $\chi ^2$ per degree of freedom.

\item Respecting the low density limit for the $p$-wave part of
  the potential.

\item Requiring theoretically acceptable values for the ratios of
real to imaginary parts of the empirical quadratic terms.

\end{enumerate}
\noindent
No {\it a priori} restrictions were placed on the $s$-wave part of the
potential as it is the topic being studied in the present work.

The final results were  obtained by usually varying only 6
or 7 parameters, as is discussed in the next section.
All fits were repeated for the four sets of density distributions mentioned
in the previous section, namely for two sets of neutron radii using
MAC densities and for the same two sets of neutron radii using SP densities.
The whole process was repeated six times, using the conventional (C)
potential, the Weise (W) potential, two RIA potentials with different
values of the parameter $a$ (Eq.(\ref{equ:Bir})) and two potentials where
both the RIA and the W mechanisms were included.

\section{Results}  \label{sec:res}

Starting with the conventional potential it was easy to obtain fits with
$\chi ^2$/F, the $\chi ^2$ per degree of freedom, of about 1.9. This
represents very good fits to pionic atom data from $^{16}$O to $^{238}$U,
covering atomic states from 1$s$ to  4$f$. 
The fits to the deeply
bound 1$s$ and 2$p$ states in $^{205}$Pb were  
better than the average. In an earlier work \cite{FGa98} full consistency
was found between
 deeply bound states in $^{207}$Pb and normal 
pionic atom states. It means that, so far, the deeply
bound  states do not have any special weight in determining
the parameters of the pion-nucleus potential. 
The results 
of these large scale fits are in general agreement with previous
works \cite{BFG97,KLT90}. 
The EELL parameter $\xi$ was not well determined
and a shallow minimum was obtained for values around 2. However,
the scattering volume parameters $c_0$ and $c_1$ turned out to be close
to the free pion-nucleon values only when $\xi$ was close to 1 and we 
 chose to keep the
value of $\xi$ fixed at 1. The parameter $c_1$ was then found to be consistent
with the free pion-nucleon value and consequently it was kept fixed at that
value in the final fits. The parameter $c_0$ too was kept fixed 
at the free pion-nucleon value in part
of the fits, as is shown in Table \ref{tab:res}, 
although marginal improvements in the fits could be obtained when
it was varied slightly. 
Comparing the present results with previous fits, we note that in
Ref.\cite{BFG97} we showed results for $\xi$=0 because that was
the {\it only} way to compare with Ref.\cite{KLT90}, due to the different
structure of the EELL effect in the two approaches. However, the results for
$\xi$=0 were not the best possible. Nevertheless, in order to check the
sensitivity of our conclusions regarding the $s$-wave potential to the
$p$-wave potentials used, we show in Table \ref{tab:resprime} results for
$c_0, c_1$ and $\xi$ fixed at the values of \cite{BFG97}.

Turning to the present work,
each fit was repeated 8 times: for the two sets
of  rms radii for neutron density distributions for each of the MAC and SP 
models used to generate the density distributions and  then for
$c_0$ fixed or  variable. All the fits 
for the conventional (C) potential produced almost the same
values of $\chi ^2$/F which makes it impossible to prefer any one of them
on the basis of quality of fit. Introducing additional absorption terms
proportional to $\rho _p^2$ \cite{KLT90} produced insignificant improvements,
e.g. $\chi ^2$ changed by less than 1.9 (out of 200) with the addition of
such a term in the $p$-wave part and by 0.1 with the addition of such a term
in the $s$-wave part.
One difference compared to the earlier analysis \cite{Fri02} was that 
Re$C_0$ was not consistent with zero and therefore it was varied in all fits.

The other five potentials, each with
additional dependence on the nuclear density
relative to the conventional 
potential, were used in  fits to the data  in much the same way.
Typical results are summarized in Table \ref{tab:res}. 
In this table the Weise
model (Eq.(\ref{equ:ddb1})) is labelled by W and the two RIA models with
$a$=2.7 and $a$=1.56 fm$^3$ (Eq.(\ref{equ:Bir})) are labelled by B1 
and B2, respectively. The rows WB1 and WB2 are for fits when 
the Weise and the RIA models were combined. It is seen from the table that
the quality of the fits is essentially the same for all the potentials 
and therefore
other criteria must be applied in order to prefer one potential or the
other.

As was stated above, the present work is focused on the empirical
values of $b_0$ and $b_1$ and their relation to the corresponding values for
the free pion-nucleon interaction. Because the parameter $b_0$ is
extremely small, we assess the various potentials essentially by the
empirical values of $b_1$ with emphasis on the effects due to the
Weise prescription (Eq.(\ref{eq:fpi2})). In this context it is particularly
important to mention the weak coupling between the $s$-wave part and
the $p$-wave part of the best-fit potentials.
This is demonstrated by comparing the results in Table \ref{tab:res}
with those in Table \ref{tab:resprime}, where the latter contains results of
similar calculations as in the former but with the linear $p$-wave
part of the potential taken from Ref.\cite{BFG97}. All three parameters
($c_0, c_1, \xi$) are very different in  the two sets of calculations
yet the corresponding empirical parameters of the $s$-wave part
are essentially the same.
Note in particular the corresponding values of $b_1$ and their uncertainties.
Such small uncertainties can be achieved {\it only} in large scale
fits and cannot possibly be achieved when studying isolated pieces
of data \cite{IOH00,KYa01}.  
The variation of $b_1$ between the different models does not
depend on the choice of the $p$-wave part, provided best fit to the data
has been achieved. Comparing the ratios Re$C_0$/Im$C_0$ in the two
tables, Table \ref{tab:res} is preferred. In what follows we will
refer only to the results where the $p$-wave part respects the
low density limit, with $\xi$=1, as in Table  \ref{tab:res}.

The anomalous repulsion in the $s$-wave part of the potential can be
seen in Table \ref{tab:res} through values of $b_1$ and Re$B_0$.
For the C potential the former is significantly more repulsive than
the free pion-nucleon value and the latter, which is the dispersive
part of the absorptive term Im$B_0$, is repulsive with a magnitude  3-4
times larger than the absorption. This is quite unacceptable as various
theoretical approaches \cite{GOS88,OGN95,KRi78,SHO95}
suggest that the real part must have about the same
magnitude as the imaginary part.
The other rows in the table show various degrees of reduced repulsion in 
$b_1$ and in Re$B_0$. This is achieved through the density dependence of $b_1$
in the Weise model which makes $b_1$ progressively more repulsive with
increasing  nuclear density and through the repulsion generated by the RIA
term. In both cases the phenomenological repulsion required by the data
is obtained while keeping $b_1$ closer to the free nucleon value and
keeping Re$B_0$ closer to expectations.

\begin{figure}
\epsfig{file=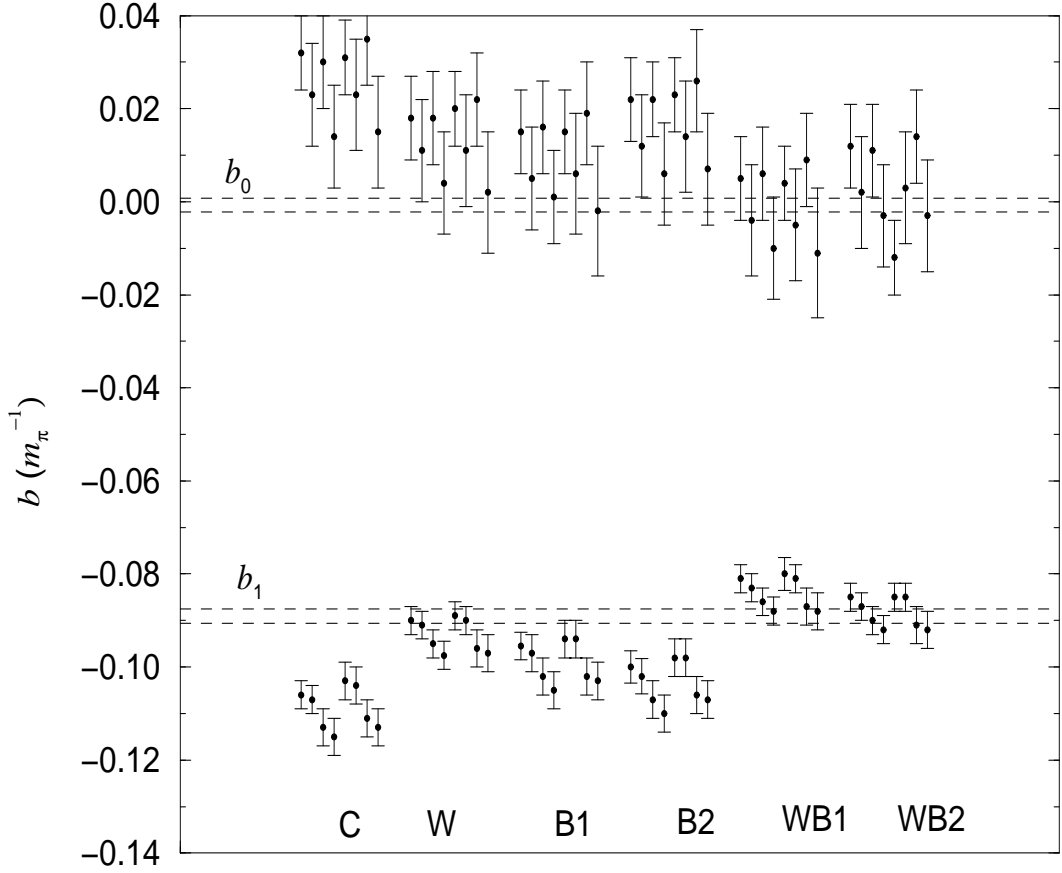, height=120mm,width=140mm}
\caption{Values of $b_0$ and $b_1$ obtained from fits to pionic atom data
for the six potentials indicated and the various densities and fit
procedures (see text). The horizontal bands indicate the experimental
values for the free pion-nucleon interaction
\protect \cite{SBG01}.}
\label{fig:b01}
\end{figure}

\begin{figure}
\epsfig{file=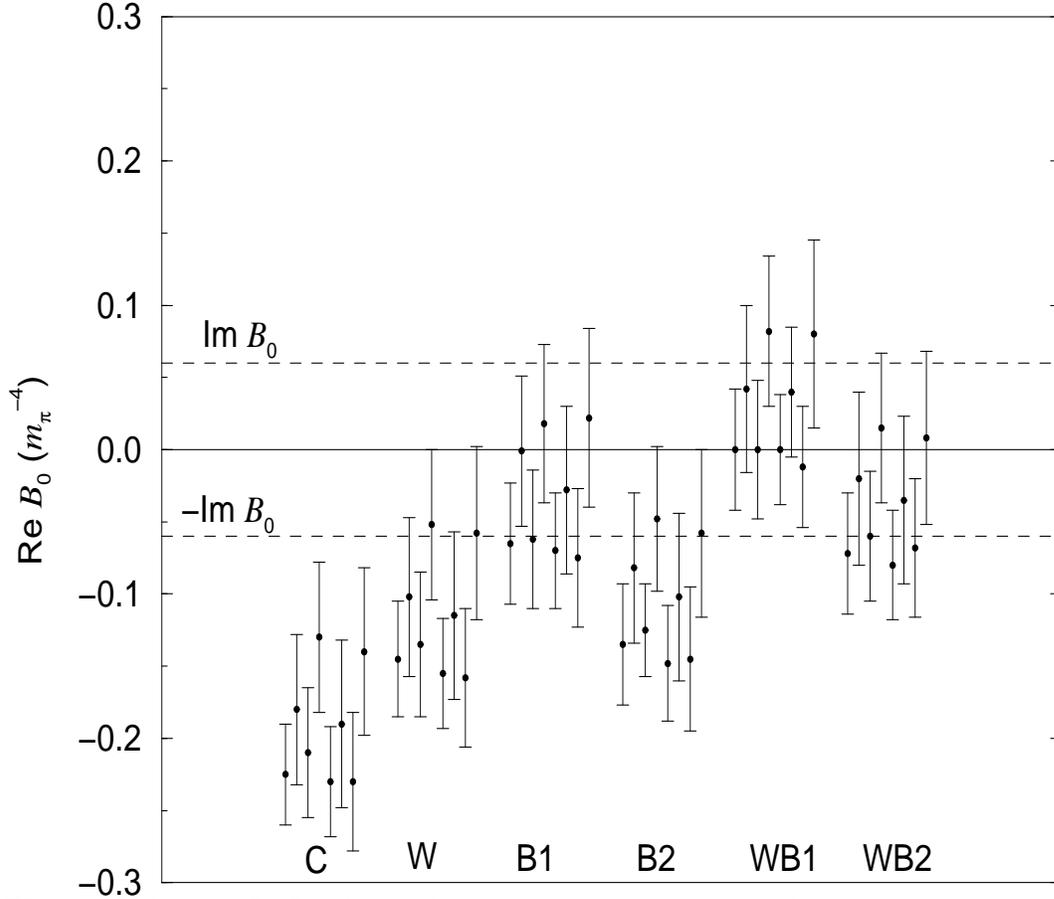, height=120mm,width=140mm}
\caption{Values of Re$B_0$ obtained from fits to pionic atom data
for the six potentials indicated and the various densities and fit
procedures (see text). The horizontal bands indicate the theoretically
expected range of values.}
\label{fig:B}
\end{figure}

Figures \ref{fig:b01} and \ref{fig:B} show the values of $b_0$, $b_1$ and 
Re$B_0$ for all the 48 different fits. The six groups 
of eight points each are according to
the  potential (see Table \ref{tab:res}), the type of nuclear 
density used and for both fixed $c_0$ and variable $c_0$. Figure \ref{fig:b01}
includes, between  horizontal dashed lines, 
the experimental values of the free
pion-nucleon $b_0$ and $b_1$ parameters. 
It is seen that the WB1 and WB2 models yield the best agreement
with experiment. The horizontal 
dashed lines in Figure \ref{fig:B} show the range
of expected values of Re$B_0$ between $-$Im$B_0$ and Im$B_0$. 
Within each
potential one can see a fairly small dependence on details such as
type of density and whether $c_0$ was varied or not. The anomalous
repulsion is clearly seen for the conventional potential. It essentially
disappears for potentials WB1 and WB2. 

\begin{figure}
\epsfig{file=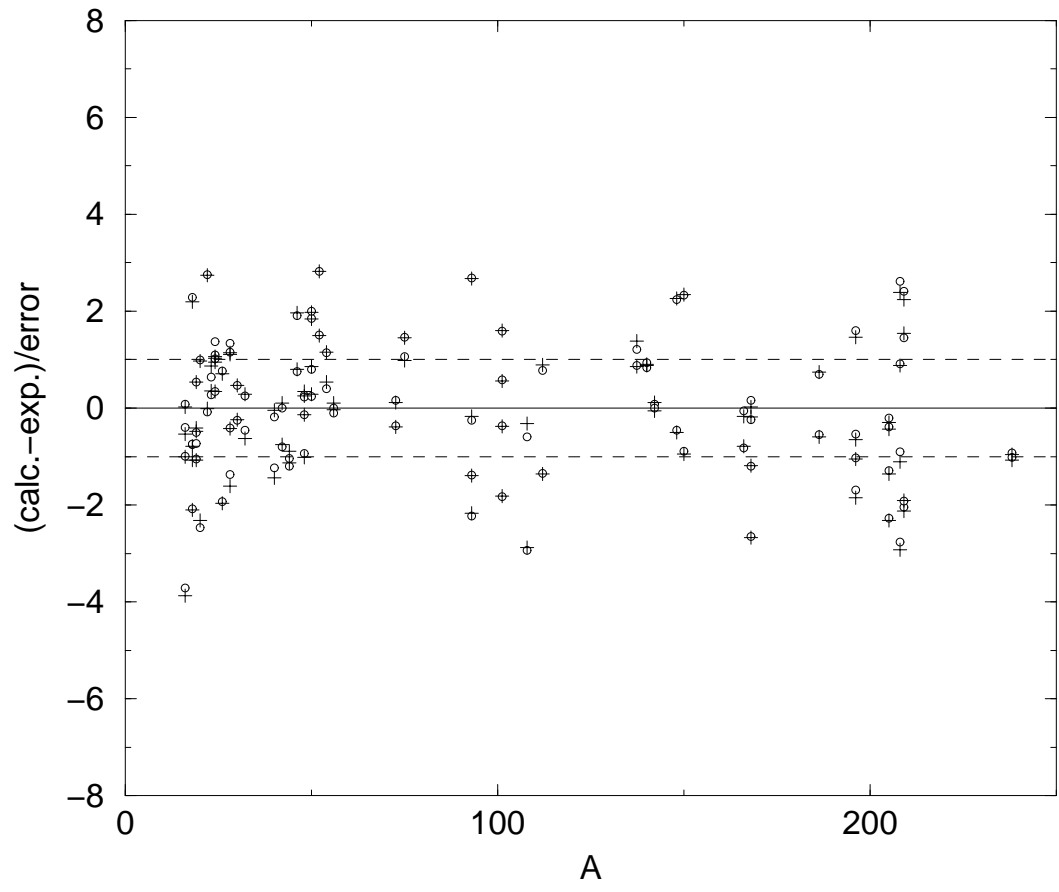, height=120mm,width=140mm}
\caption{Values of $\chi$ for the best fit C potential (open circles)
and W potential (+ signs) using MAC densities.}
\label{fig:chi}
\end{figure}

Figure \ref{fig:chi} shows values of $\chi$ for all the data points, for
fits based on the C and W potentials with MAC densities. 
It illustrates  that there are no systematic differences
between the different potentials from the point of view of fits to the data.
 Very similar
results are obtained when different nuclear densities are used. It seems 
that whatever systematics is observed in Fig. \ref{fig:chi} it 
is peculiar to the targets and not to differences between the models
used here. Obviously it could indicate deficiencies in the models.

\begin{figure}
\epsfig{file=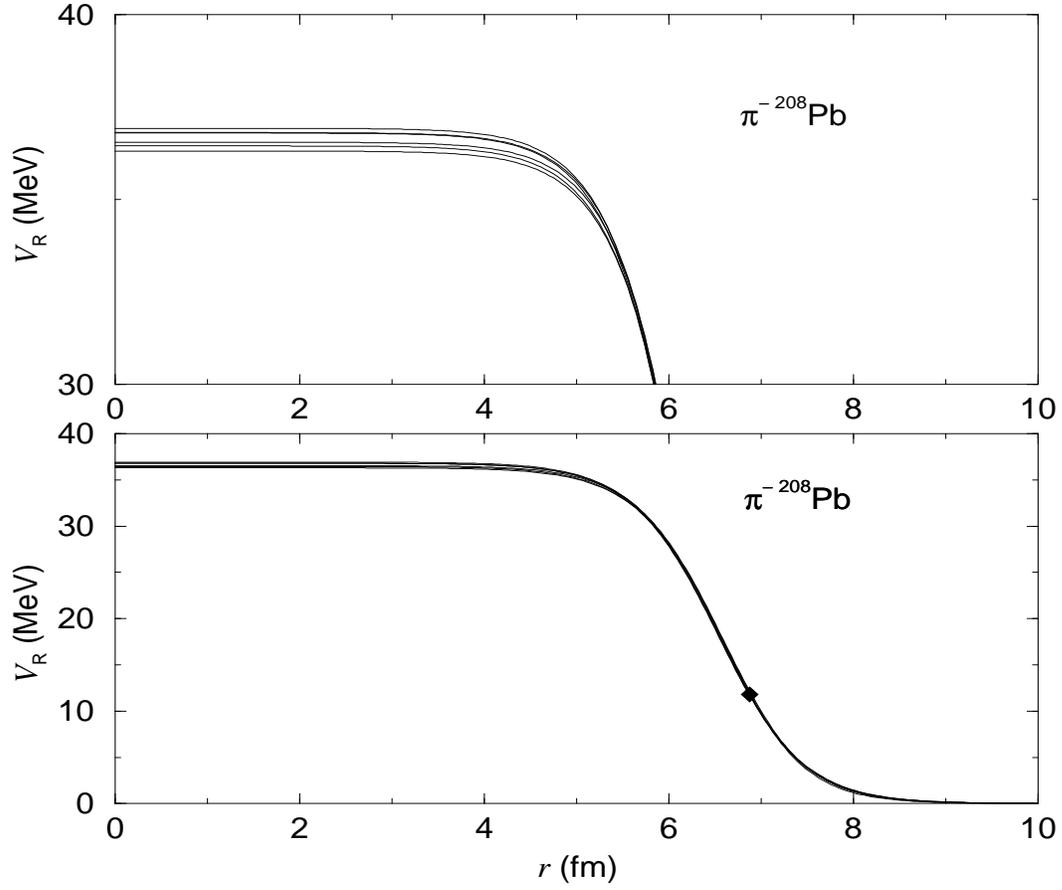, height=120mm,width=140mm}
\caption{The real part of the $s$-wave  
$\pi ^-$ $^{208}$Pb potential for the six models using MAC densities.
The `half-density' point is indicated at 6.9 fm.}
\label{fig:potl1}
\end{figure}

\begin{figure}
\epsfig{file=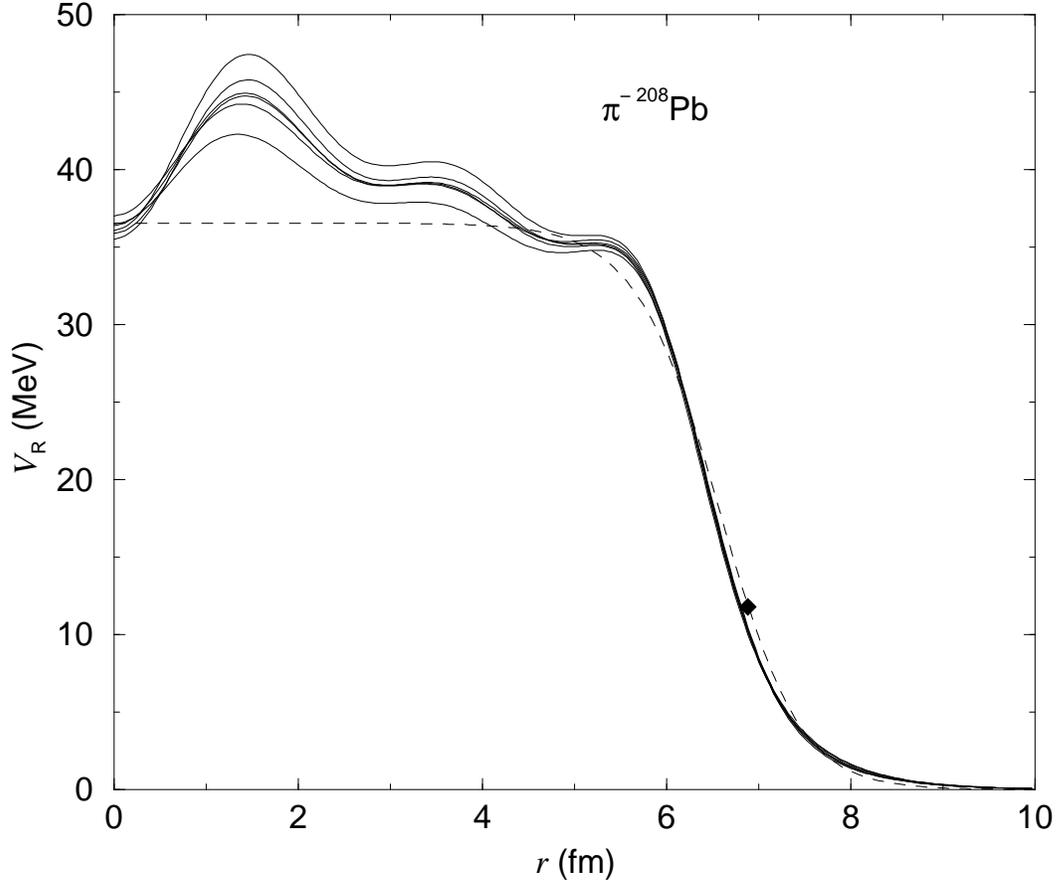, height=120mm,width=140mm}
\caption{The real part of the $s$-wave 
$\pi ^-$ $^{208}$Pb potential for the six models using SP densities. 
Also shown (dashed)
is the C potential based on MAC densities. The `half-density' point
is indicated at 6.9 fm.}
\label{fig:potl2}
\end{figure}

Another topic of interest is the way  the real $s$-wave potential 
varies from the nuclear
surface towards the nuclear interior. As an example Fig. \ref{fig:potl1}
shows the real part of the  $\pi ^-$ $^{208}$Pb potential
 for the six models
of the present work, based on MAC densities. All six potentials produce
almost identical fits to the data (see Table \ref{tab:res}) and indeed
they are almost identical throughout the nucleus. Note that the potentials
do {\it not} follow the nuclear density distribution because of the various
non-linear terms. The almost unique potential all
over the nucleus is a result of the featureless shape of the MAC densities.
Figure \ref{fig:potl2} shows, again for $^{208}$Pb,  the six potentials
this time based on SP densities. The C potential for MAC densities is
also included as a dashed line. Here we see that all the potentials 
essentially agree in the surface region near 6.2 fm, 
which presumably is the region best determined by the data. 
The potential at that point is about 25 MeV whereas 
the point where the density is half of the central density 
(shown on the figures) is at 6.9 fm
and the potential there is about 11.8 MeV.
The potentials due to the various models, based on the SP densities, differ
in the interior because  these  models  extrapolate differently
into the nuclear interior due to the structure of the densities and the
variety of non linear effects. It is probably safe to conclude that the
real potential in the nuclear interior is close to 30-35 MeV, in contrast
with smaller values  \cite{YHI98,IOH00} advocated earlier but in
reasonable agreement with  the value of 
28$\pm$3 MeV of Friedman and Gal \cite{FGa98} and with 
the latest result of Geissel et al \cite{GGG02}.

\section{Summary, conclusions and outlook}  \label{sec:sum}

The prime motivation behind the present work was the recent suggestion
\cite{Wei01,KWe01} that the anomalous $s$-wave repulsion of pions
in nuclear matter has its origins in a density dependence of the
pion decay constant which reflects the change of QCD vacuum structure
in dense matter. The consequences for pionic atoms are evident then 
through the isovector parameter $b_1$ of the
potential, which is quite important because of the extremely small value
of the isoscalar parameter $b_0$. The density dependence of the pion
decay constant (Eq.(\ref{eq:fpi2})) 
leads to an additional well-defined density dependence in the 
pion-nucleus potential which can be tested by fits to pionic atom data.
The idea was to check the consequences of prescription (\ref{eq:fpi2})
by applying it to modern and extensive pionic atom data. Possible
modifications of $B_0$ have not been considered because we treat this parameter
all along as purely phenomenological without any experimental or theoretical
reference values, except for the expectation that $|$Re$B_0|$ cannot exceed
much the value of Im$B_0$.
Also included was a RIA term, which 
in some sense is equivalent to imparting a density dependence to the
underlying nucleon mass and
was shown \cite{GJF92} to explain
part of the repulsion, although this term is not unique.
A comment on this renormalization and on that of $f_\pi$ (Eq.(\ref{eq:fpi2}))
has been made recently by Brown and Rho \cite{BRo02}.
A very broad data base had been used and extra care was 
taken in choosing the nuclear densities which
are essential ingredients of the potential. Radii of neutron
distributions were varied slightly, and were constrained by 
recent RMF calculations. If one assumes that radii of neutron density
distributions are equal to the corresponding radii for proton distributions,
then the fits become totally unacceptable.

The results of the present work show that all the potentials produce
rather equivalent fits to the data, displaying small sensitivity
to the type of potential, to the type of density or to the precise values
of the rms radii of the neutron distributions. The differences between
the various potentials are mostly in the value of $b_1$ and how
close it is to the free pion-nucleon value. The prescription
Eq.(\ref{equ:ddb1}) removes most of the `anomaly' and when an RIA term
is also included then the parameters $b_0$, $b_1$ and Re$B_0$ have
most acceptable values. These conclusions provide support  to the validity
of the correction  suggested by Weise
 \cite{Wei01} to the conventional pion-nucleus interaction.   

Finally, it is interesting to study the above mentioned 
features at energies just above
threshold through the elastic scattering of  low
energy pions by nuclei, thus testing further the validity of the chirally
motivated approach \cite{Wei01}.
Indeed it has been shown
\cite{SMC79,SMa83,MFJ89} that  pion-nucleus
potentials  develop smoothly from the bound states regime to the
elastic scattering regime. An experiment to measure the elastic scattering
of 20 MeV $\pi ^{\pm}$ by several nuclei with that aim in mind has been
approved recently at the Paul Scherrer Institute (PSI) \cite{FBC02}.

\vspace{1cm}
I wish to acknowledge  many stimulating discussions with   A. Gal.

\vspace{1cm}
This research was partially supported by the Israel Science Foundation.

\begin{table}
\caption{Parameter values from fits to 106 pionic atom data points. Other
$p$-wave parameters were held fixed at $c_0$=0.22$m_\pi^{-3}$,
$c_1$=0.18$m_\pi^{-3}$ and $\xi$=1. The free pion-nucleon values 
\protect \cite{SBG01}
are $b_0=-0.0001^{+0.0009}_{-0.0021} m_\pi ^{-1}$ and
$b_1=-0.0885^{+0.0010}_{-0.0021} m_\pi ^{-1}$}
\label{tab:res}
\begin{tabular}{lccccccc}
potl.&$\chi ^2$/F&$b_0$ ($m_\pi^{-1}$)&$b_1$ ($m_\pi^{-1}$)&
 Re$B_0$ ($m_\pi^{-4}$)&Im$B_0$($m_\pi^{-4}$)&
Re$C_0$ ($m_\pi^{-6}$)&  Im$C_0$($m_\pi^{-6}$) \\
\hline
C & 1.94 & 0.030$\pm$0.010&$-$0.113$\pm$0.004&$-$0.21$\pm$0.04&0.055$\pm$0.002&
$-$0.029$\pm$0.009&  0.064$\pm$0.004 \\
W & 1.93 & 0.018$\pm$0.010&$-$0.095$\pm$0.003&$-$0.14$\pm$0.05&0.054$\pm$0.002&
$-$0.026$\pm$0.009&  0.063$\pm$0.004 \\
B1&1.95 & 0.016$\pm$0.010&$-$0.102$\pm$0.004&$-$0.06$\pm$0.05&0.054$\pm$0.002&
$-$0.026$\pm$0.009 &   0.064$\pm$0.004 \\
B2&1.94&0.022$\pm$0.008&$-$0.107$\pm$0.004&$-$0.12$\pm$0.03&0.054$\pm$0.002&
$-$0.028$\pm$0.008 & 0.064$\pm$0.003 \\
WB1&1.96&0.006$\pm$0.010&$-$0.086$\pm$0.003&0.00$\pm$0.05&0.053$\pm$0.002&
$-$0.024$\pm$0.009 & 0.063$\pm$0.003 \\
WB2&1.94&0.011$\pm$0.010&$-$0.090$\pm$0.003&$-$0.06$\pm$0.04&0.054$\pm$0.002&
$-$0.025$\pm$0.009 & 0.063$\pm$0.003 \\
\end{tabular}
\end{table}

\begin{table}
\caption{Parameter values from fits to 106 pionic atom data points. Other
$p$-wave parameters were held fixed at $c_0$=0.261$m_\pi^{-3}$,
$c_1$=0.104$m_\pi^{-3}$ and $\xi$=0. The free pion-nucleon values
\protect \cite{SBG01}
are $b_0=-0.0001^{+0.0009}_{-0.0021} m_\pi ^{-1}$ and
$b_1=-0.0885^{+0.0010}_{-0.0021} m_\pi ^{-1}$}
\label{tab:resprime}
\begin{tabular}{lccccccc}
potl.&$\chi ^2$/F&$b_0$ ($m_\pi^{-1}$)&$b_1$ ($m_\pi^{-1}$)&
 Re$B_0$ ($m_\pi^{-4}$)&Im$B_0$($m_\pi^{-4}$)&
Re$C_0$ ($m_\pi^{-6}$)&  Im$C_0$($m_\pi^{-6}$) \\
\hline
C'& 1.93 & 0.020$\pm$0.009&$-$0.114$\pm$0.004&$-$0.15$\pm$0.04&0.054$\pm$0.002&
$-$0.28$\pm$0.01&  0.062$\pm$0.003 \\
W'& 1.90 & 0.008$\pm$0.009&$-$0.097$\pm$0.004&$-$0.07$\pm$0.04&0.053$\pm$0.002&
$-$0.28$\pm$0.01&  0.067$\pm$0.003 \\
B1'&1.94 & 0.006$\pm$0.009&$-$0.104$\pm$0.004&0.00$\pm$0.04&0.053$\pm$0.002&
$-$0.28$\pm$0.01 &   0.066$\pm$0.003 \\
B2'&1.93&0.011$\pm$0.009&$-$0.108$\pm$0.004&$-$0.07$\pm$0.04&0.054$\pm$0.002&
$-$0.28$\pm$0.01 & 0.067$\pm$0.003 \\
WB1'&1.94&$-$0.005$\pm$0.009&$-$0.088$\pm$0.003&0.06$\pm$0.04&0.053$\pm$0.002&
$-$0.27$\pm$0.01 & 0.066$\pm$0.003 \\
WB2'&1.91&0.001$\pm$0.009&$-$0.092$\pm$0.003&0.00$\pm$0.04&0.053$\pm$0.002&
$-$0.28$\pm$0.01 & 0.066$\pm$0.003 \\
\end{tabular}
\end{table}

\end{document}